\documentclass[journal,onecolumn]{IEEEtran}
\usepackage[dvipsnames]{xcolor}
\newcommand{\CM}[1]{\textcolor{Black}{#1}}

\usepackage{generic}
\usepackage{cite}
\usepackage{amsmath,amssymb,amsfonts}
\usepackage{algorithmic}
\usepackage{graphicx}
\usepackage{textcomp}

\def\BibTeX{{\rm B\kern-.05em{\sc i\kern-.025em b}\kern-.08em

T\kern-.1667em\lower.7ex\hbox{E}\kern-.125emX}}

\begin{document}

\title{Influence of Carrier-Carrier Interactions on the Sub-threshold Swing of Band-to-Band Tunnelling Transistors\\
}
\author{Chen Hao Xia,
Leonard Deuschle, Jiang Cao, Alexander Maeder, and
Mathieu Luisier
\thanks{This research was supported by NCCR MARVEL, funded by the Swiss National Science Foundation (SNSF) under grant No. 205602, and by SNSF under grant No. 209358 (QuaTrEx). We acknowledge support from CSCS under project s1119.}
\thanks{The authors are with the Integrated Systems Laboratory, ETH Zurich, 8092 Zürich, Switzerland (e-mail: chexia@student.ethz.ch).}}
\maketitle

\begin{abstract}
Band-to-band tunnelling field-effect transistors (TFETs) have long been considered as promising candidates for future low-power logic applications. However, fabricated TFETs rarely reach sub-60 mV/dec sub-threshold swings (SS) at room temperature. Previous theoretical studies identified Auger processes as possible mechanisms for the observed degradation of SS. Through first-principles quantum transport simulations incorporating carrier-carrier interactions within the Non-equilibrium Green's Function formalism and self-consistent GW approximation, we confirm here that Auger processes are indeed at least partly responsible for the poor performance of TFETs. Using a carbon nanotube TFET as testbed, we show that carrier-carrier scattering alone significantly increases the OFF-state current of these devices, thus worsening their sub-threshold behavior. 

\end{abstract}


\section{Introduction}

Unlike metal–oxide–semiconductor field-effect transistors (MOSFETs), band-to-band tunnelling FETs (TFETs) operate through the injection of cold instead of hot carriers and are thus theoretically capable of providing a sub-threshold swing (SS) below 60~mV/dec at room temperature. \cite{peng2014carbon,das2021beyondcmos}. \CM{However,  very few fabricated TFETs display sub-thermionic switching over a large current window while retaining the high ON-state current needed to achieve ultra low power consumption in practical applications.} Only few experimental breakthroughs have been reported in literature, for example \cite{appenzeller2004,seabaugh2013tunneling, pang2021steep}. 

Previous theoretical studies suggested Auger processes as a possible explanation for SS degradation in experimental TFETs \cite{teherani2016auger,ahmed2020comprehensive}. It was postulated that in the sub-threshold region, Auger processes open up additional tunnelling paths at the source-channel junctions, which increases the OFF-state current and imposes a lower limit on the SS of TFETs. \CM{This hypothesis has not yet been verified through device simulation because it is computationally challenging to account for Auger processes in quantum transport calculations, which are required to investigate TFETs.} Hence, we perform in this paper \textit{ab initio} simulations of TFET models based on the Non-equilibrium Green's function (NEGF) formalism, including electron-electron (e-e) interactions within the self-consistent GW (scGW) approximation. \CM{Carbon nanotube (CNT) devices are chosen as testbeds because of their limited computational times, even in the context of NEGF+scGW, and their excellent band-to-band tunnelling characteristics \cite{koswatta2005computational,Shirazi2019gaacnt}. Furthermore, the design of CNT-based TFETs offers various degrees of freedom that allow it to mimic the behaviour of other types of TFETs and generalize the conclusions of this study.}

\begin{figure}
    \centering
    \includegraphics[width=1\linewidth]{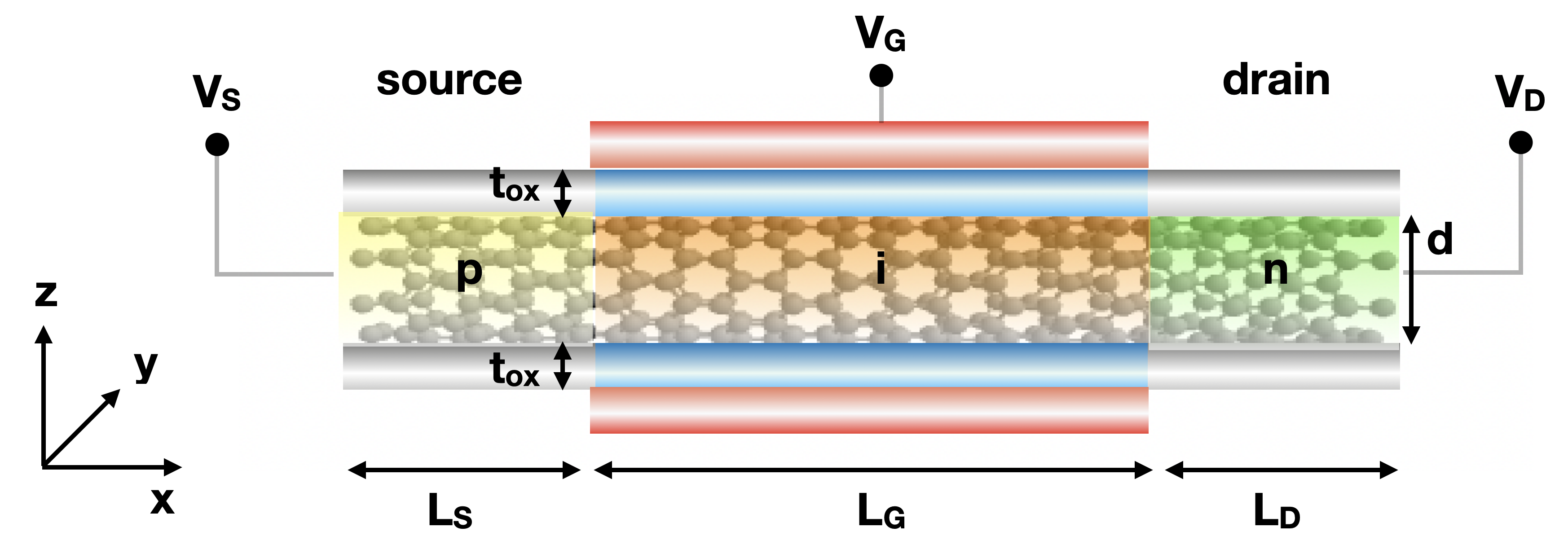}
    \caption{Schematic of a gate-all-around CNT TFET structure made of 4608 atoms. It consists of a single-wall CNT with diameter d = 0.72 nm and gate length $L_G$ = 21.6 nm, surrounded by an oxide layer of thickness $t_{ox}$ = 3 nm and relative permittivity $\varepsilon_R$ = 20. Source and drain are 20 nm long ($L_S$ = $L_D$ = 20 nm). They are $p$- ($4.6\cdot10^{-3}$  holes per atom)  and $n$-doped ($2.1\cdot10^{-3}$ electrons per atom), respectively. \CM{Spacers with $\varepsilon_R$ = 3.9 surround the source and drain}. A voltage $V_G$ is applied to the gate, $V_D$ to the drain, with source grounded.}
    \label{fig:tfetstructure}
\end{figure}

\CM{Our baseline TFET structure used throughout the paper is shown in Fig.~\ref{fig:tfetstructure}. Note that to isolate the impact of e-e interactions, all other scattering sources, e.g., electron-phonon or interface traps, have been excluded as they have been extensively studied before \cite{koswatta2005computational,Schenk}. We aim to demonstrate that e-e interactions alone can partly explain the relatively poor performance of experimental TFETs. After describing the NEGF+scGW methodology, we will therefore highlight the importance of Auger processes in TFETs.}

\section{Method}
\begin{figure}[t]
    \centering
    \includegraphics[width=0.9\linewidth]{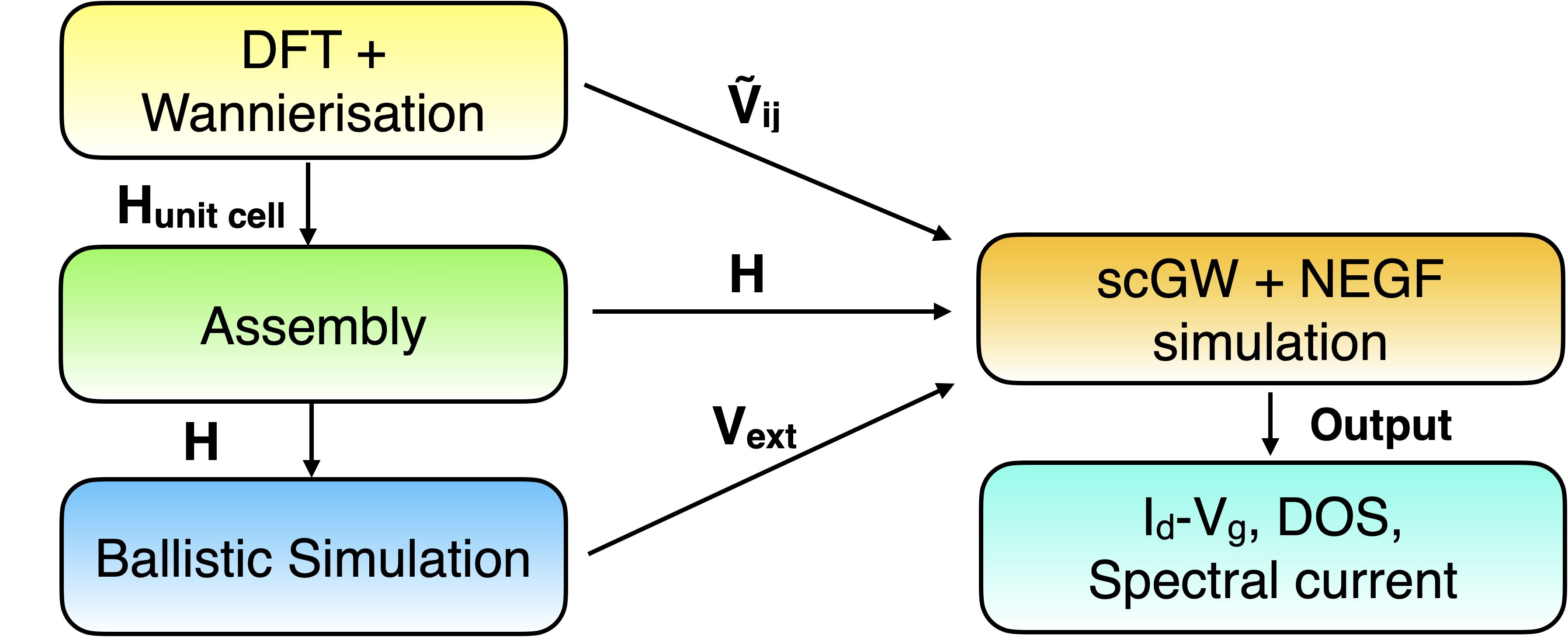}
    \caption{Simulation workflow. The Coulomb matrix $\widetilde{V}_{ij}$, device Hamiltonian $H$ and electrostatic potential $V_{ext}$ form the input parameters to the developed NEGF + scGW solver. }
    \label{fig:flowchart}
\end{figure}

The simulation workflow to capture e-e scattering at the \textit{ab initio} level is outlined in Fig.~\ref{fig:flowchart}. It starts with a DFT calculation of the selected (8,0) single-wall unit cell with Quantum Espresso \cite{giannozzi2009quantum} and GGA-PBE pseudo-potentials \cite{prandini2018precision}. The resulting plane-wave quantities are then transformed into a set of maximally localised Wannier functions (MLWF) with the wannier90 tool \cite{mostofi2008wannier90}, keeping 32 orbitals per unit cell, one for each atom. The PBE band gap of the (8,0) CNT is equal to 0.54 eV. It is adjusted to 0.8 eV to ensure optimal ballistic performance according to \cite{zhang2014optimum}, by applying a scissor operator to the unit cell Hamiltonian $H_{WF,C}$. \CM{This operation consists of transforming $H_{WF,C}$ back to its $k$-space representation $H_k$, followed by a diagonalisation to obtain}
\begin{equation}
\label{eq:scissors}
    H_k = U_kDU_k^{\dagger},
\end{equation}
\CM{where the columns of ${U_{k}}$ contain the rotated Bloch vectors and $D$ is a diagonal matrix representing the eigen-energies of the system. To increase the band gap, all entries of $D$ corresponding to either the conduction or valence band are shifted up or down in energy, respectively. $H_k$ is then transformed into the modified unit cell Hamiltonian $\tilde{H}_{WF,C}$ and assembled to form the full device Hamiltonian with increased band gap.} 

In the NEGF formalism, the following equation must be solved to obtain the retarded Green's function $G^R$ 
\begin{equation}
\left(E\cdot I-H-V_{ext}-\Sigma^{R}_{B}(E)- \Sigma^{R}_{GW}(E)\right) \cdot G^{R}(E)  = I,
\label{eq:GR}
\end{equation}
where $E$ is the electron energy, $I$ the identity matrix, $H$ the Hamiltonian matrix of the device, and $V_{ext}$  the electrostatic potential ,while $\Sigma^{R}_{B}$ and $\Sigma^{R}_{GW}$ are the retarded boundary and GW self-energies, respectively \cite{datta}. The lesser and greater Green's function $G^{\lessgtr}$ are given by 
\begin{equation}
\label{eq:lessgtrGLGtilda}
   G^{\lessgtr}(E) = G^R(E)\cdot \Sigma^{\lessgtr}(E) \cdot {G^{R}}^{\dagger}(E).
\end{equation}
$\Sigma^{\lessgtr}$ denotes the sum of the boundary $\Sigma^{\lessgtr B}$  and GW $\Sigma^{\lessgtr GW}$ self-energies  . Initially, all $\Sigma_{GW}$ are set to 0 and a ballistic simulation of the TFET is performed  self-consistently with Poisson's equation \cite{luisier2008omen}. The same electrostatic potential as in the ballistic case is then used when e-e interactions are taken into account to reduce computational intensity. The lesser/greater GW self-energies are expressed as 
\begin{equation}
\label{eq:siglg}
\Sigma^{\lessgtr}_{GW,ij} = i \int dE' \lbrack G^{\lessgtr}_{ij}(E')W^{\lessgtr}_{ij}(E-E') \rbrack,
\end{equation}
where $W$ is the screened Coulomb interaction \cite{paper:hedin,paper:thygesen}. The indices $i$ and $j$ represent the orbital indices of atoms situated at positions $\mathrm{\mathbf{R}_i}$  and $\mathrm{\mathbf{R}_j}$,  respectively. Interaction distances satisfying $|\mathrm{\mathbf{R}_i}-\mathrm{\mathbf{R}_j}|\leq0.64$ nm are kept in Eq.~(\ref{eq:siglg}) . The retarded GW self-energy can be split into an exchange ($\Sigma^{R}_{x}$) and correlation ($\Sigma^{R}_{corr}$) component

\begin{equation}
\label{eq:sigr_split_energydomain}
\Sigma^{R}_{GW, ij}(E) = \Sigma^{R}_{x} + \Sigma^{R}_{corr}(E),
\end{equation}

where

\begin{equation}
\label{eq:sigr_x_ed}
\Sigma^{R}_{x, ij} = i\int_{E}dE \text{} G^{<}_{ij}(E)\widetilde{V}_{ij}.
\end{equation}
$\widetilde{V}_{ij}$ is the bare Coulomb interaction between orbitals $i$ and $j$,
\begin{equation}
\label{eq:PI}
\Sigma^{R}_{corr, ij}(E) = - \frac{i}{2}\Gamma_{ij}(E) + \mathcal{P} \text{ } \int_{E} \frac{dE'}{2\pi}\frac{\Gamma_{ij}(E')}{E-E'}.
\end{equation}
In Eq. \eqref{eq:PI} $\mathcal{P}$ denotes the principal part of the integral and $\Gamma_{ij}$ is the broadening function defined as \cite{Lake1997}
\begin{equation}
    \Gamma_{ij} = i \lbrack \Sigma^{>}_{GW,ij} - \Sigma^{<}_{GW,ij} \rbrack.
\end{equation}

The energy convolutions in Eq. (\ref{eq:siglg}) are calculated in the time domain through Fourier transforms of all variables. Only the imaginary part of Eq.~\eqref{eq:sigr_split_energydomain} is retained to fix the band gap of the device and allow for direct comparison with ballistic simulations. The retarded, lesser, and greater screened interactions $W$ and irreducible polarisation $P$  are calculated via transformed Hedin's equations \cite{deuschle2023ab,cao2023ab}. The bare Coulomb matrix elements $\widetilde{V}_{ij}$ are computed according to \cite{paper:thygesen}. Equations.~\eqref{eq:GR} and \eqref{eq:lessgtrGLGtilda} are solved self-consistently with Hedin's equations for $W$ and $P$. Upon convergence, the density-of-states (DOS) and energy-resolved electrical and energy currents can be extracted \cite{book:pourfath2014non}. Our formalism satisfies both current and energy conservation. 






\section{Results}
The proposed scheme was applied to the CNT TFET of Fig. \ref{fig:tfetstructure}. It includes a gate-all-around architecture \cite{koswatta2005computational} that  offers excellent electrostatic control. The device body is made of a (8,0) CNT, with a diameter of 7.2 $\mathrm{\AA}$ and consists of 144 unit cells, each containing 32 atoms. It is surrounded by an oxide layer with a thickness of \CM{3 nm and a relative permittivity of $\epsilon_R$=20}. The structure is 61.6 nm long, with a gate length $L_G$ = 21.6 nm and source/drain lengths $L_S$ = $L_D$ = 20 nm. These dimensions were chosen to minimize short-channel effects. Asymmetric $p$ and $n$ doping is also applied to the source and drain to  suppress channel-to-drain tunnelling in the OFF-state, with $\mathrm{N_A} = 4.6\cdot10^{-3}$ holes per atom in the source and $\mathrm{N_D} = 2.1\cdot10^{-3} \text{electrons per atom}$ in the drain. For simplicity, the doping is assumed homogeneous, with sharp edges at the source-channel and channel-drain interfaces. \CM{To reduce the influence of the gate contact on the source and drain through fringing fields, low-$\kappa$ spacers with $\epsilon_R$=3.9 cover these regions.} 

\begin{figure}
    \centering
    \includegraphics[width=1\linewidth]{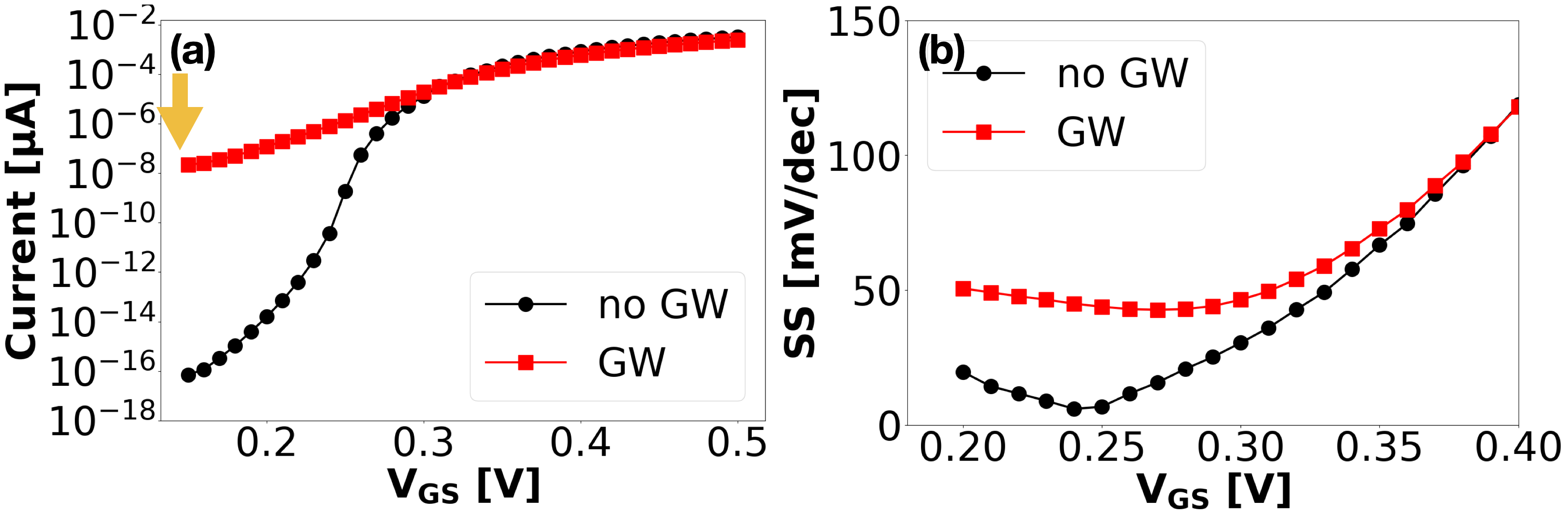}
    \caption{(a) Transfer characteristics $I_D-V_{GS}$ at $V_{DS}$=0.5 V of the CNT TFET in Fig. \ref{fig:tfetstructure} without e-e interactions (black line with circles) and with e-e interactions (red line with squares). The arrow indicates the bias point $V_{GS}$ = 0.15 V. (b) Corresponding SS as a function of $V_{GS}$.} 
    \label{fig:idvg}
\end{figure}

\begin{figure}
    \centering
    \includegraphics[width=1\linewidth]{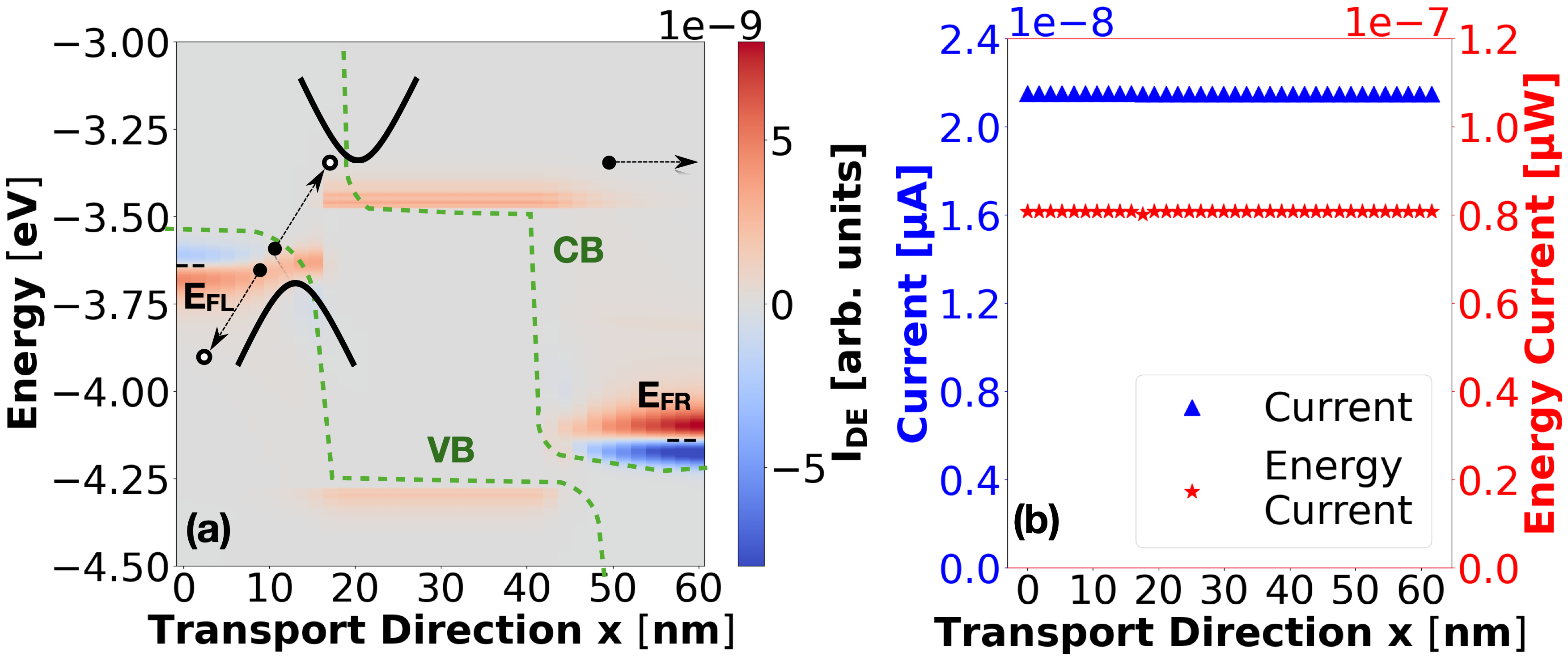}
    \caption{(a) Spectral distribution of the electron flow $I_{DE}$ along the TFET transport direction ($x$) at $V_{GS}$ = 0.15 V and $V_{DS}$ = 0.5 V. Red represents the flow from left to right, blue from right to left. The band edges are marked in green. Black schematics illustrate the Auger processes. (b) Electrical (blue triangles) and energy (red stars) currents along $x$ at $V_{GS}$ = 0.15 V and $V_{DS}$ = 0.5 V.}
    \label{fig:speccurrent}
\end{figure}

The energy range in Eq.~(\ref{eq:GR}) extends between -6 and -2 eV, with a resolution of 2 meV, to ensure convergence. The gate-to-source voltage $V_{GS}$ is swept between 0.15 V and 0.5 V, at a drain-to-source voltage $V_{DS}$=0.5 V and temperature $T$=300 K. The calculated $I_D$-$V_{GS}$ transfer characteristics and SS, with and without e-e scattering, are displayed in Fig. \ref{fig:idvg}. As can be seen in sub-plot (a), the inclusion of carrier-carrier interactions drastically increases the OFF-state current by more than two orders of magnitude. Consequently, SS goes from~10 mV/dec to about 50 mV/dec for ${V_{GS}}$ $<$ 0.2V. (Fig. \ref{fig:idvg}(b)). The difference in current \CM{with and without e-e interactions} becomes negligible for $V_{GS}>$0.3 V. To uncover the origin of this increase at low $V_{GS}$, the spectral distribution of the current at $V_{GS} $= 0.15 V is plotted in Fig. \ref{fig:speccurrent}(a).

Although the minimum of the channel conduction band (x $>$ 20 nm) is located at higher energies than the top of the source valence band (x $\leq$ 20 nm), a positive electron flow (red) resulting from Auger processes is observed. It can be explained as follows: \CM{At room temperature, incoming electrons (black circles) situated above the source Fermi energy EFL recombine with high-energy holes (white circles) in the valence band. The released energy is then transferred to neighbouring electron-hole pairs, allowing electrons close to the source valence band edge to transition through the narrow band gap of the CNT to empty states in the channel conduction band. Lower energy electrons that cannot overcome the source-channel barrier after absorption end up close to the source valence band edge before being re-absorbed by this contact, giving rise to a negative current (blue).} Similar Auger processes observed on the drain side also contribute to OFF-state current increase. It is important to realise that Auger processes conserve energy, contrary to electron-phonon interactions which can also produce current distributions similar to Fig.~\ref{fig:speccurrent}(a). The energy conservation of our NEGF+scGW scheme is shown in Fig. \ref{fig:speccurrent}(b) where the electrical and energy currents are plotted along the CNT axis. Both quantities remain constant throughout the device.   

\CM{To determine whether Auger processes generally affect TFETs, we constructed 5 other device configurations with different band gaps ($E_g$=0.6, 0.8, and 1.2 eV) and electrostatic properties. By varying the permittivity of the source drain spacers $\varepsilon_s$ from 3.9 to 20, we obtain abrupt or gradual source-channel interfaces corresponding to excellent and poor electrostatics, respectively. The OFF-state current ($I_{OFF}$) of these TFETs with ($I_{GW}$) and without GW ($I_{0}$) is summarized} \CM{in Table \ref{table:examples}. A significant increase of $I_{OFF}$ with GW is observed in all cases, with current distributions similar to Fig.~\ref{fig:speccurrent}(a), thus demonstrating the importance of Auger processes in TFETs.}

\begin{figure}[t]
    \centering
    \includegraphics[width=1\linewidth]{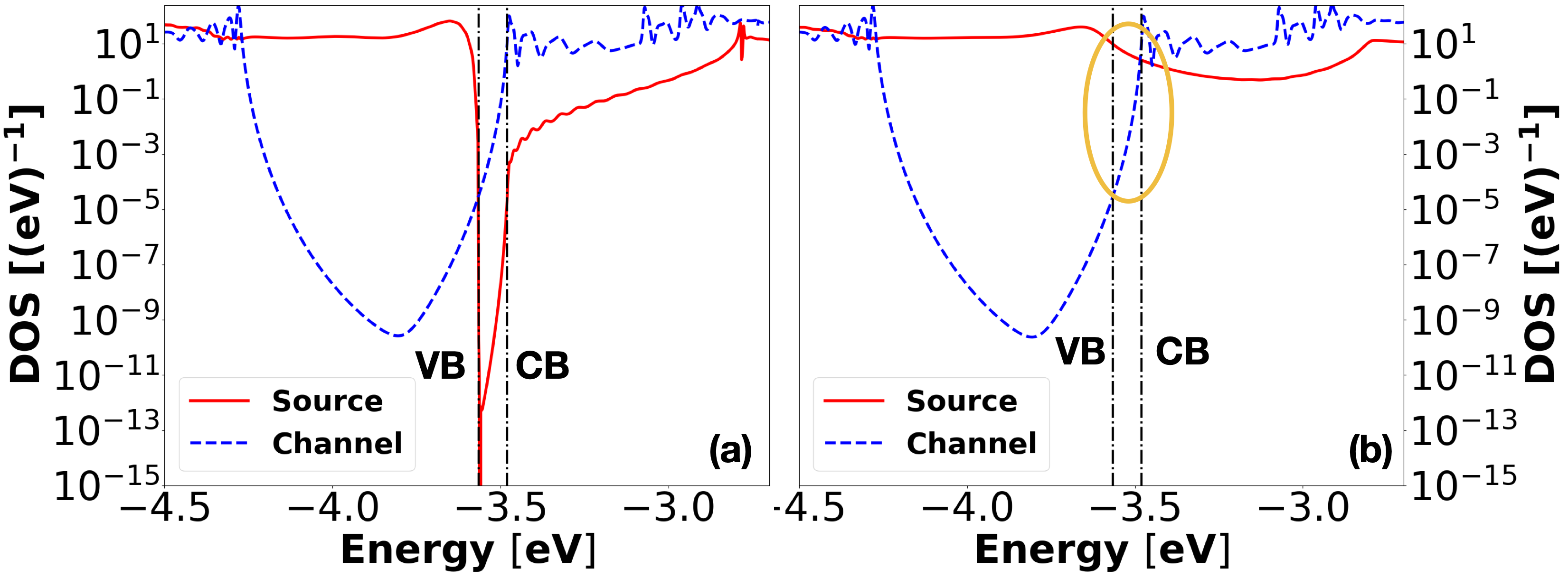}
    \caption{ (a) DOS at $x$=17.1 nm (solid red line) and 29.1 nm (dashed blue line) of the CNT TFET at $V_{GS}$ = 0.15 V and $V_{DS}$ = 0.5 V, without GW. Source valence band (VB) and channel conduction band (CB) edges are indicated by dotted black lines. (b) Same as (a), but with GW. Orange circle marks the overlap of source and channel wave functions}
    \label{fig:DOSchannel}
\end{figure}

\begin{table}
\centering
\caption{OFF-state current summary for 6 TFET examples with varying band gaps and electrostatic properties.} 
\begin{tabular}{l l l l l}
\textbf{E\textsubscript{G}(eV)}&\textbf{$\varepsilon$\textsubscript{s}}& \textbf{I\textsubscript{0}(µA)}& \textbf{I\textsubscript{GW}(µA)}& \textbf{I\textsubscript{GW}/I\textsubscript{0}}\\
\hline
0.6&3.9 & 9.24 x 10\textsuperscript{-7}& 1.13 x 10\textsuperscript{-5} & 1.2 x 10\textsuperscript{1}\\
  0.6 &20& 2.31 x 10 \textsuperscript{-10}& 1.28 x 10\textsuperscript{-7}&5.6 x 10\textsuperscript{2}\\
0.8&3.9 & 7.08 x 10\textsuperscript{-17}& 2.15 x 10\textsuperscript{-8} & 3.0 x 10\textsuperscript{8}\\
 0.8 & 20& 1.55 x 10\textsuperscript{-19}& 4.13 
 x 10\textsuperscript{-11}&2.7 x 10\textsuperscript{8}\\
1.2&3.9 & 1.51 x 10\textsuperscript{-20}& 6.58 x 10\textsuperscript{-11} & 4.3 x 10\textsuperscript{9}\\
 1.2 & 20& 5.31 x 10\textsuperscript{-21}& 1.65 x 10\textsuperscript{-13}&3.1 x 10\textsuperscript{7}\\
\end{tabular}

\label{table:examples}

\end{table}
Carrier-carrier interactions also induce a second phenomenon that increases the OFF-state current of TFETs. It can be visualized in Fig.~\ref{fig:DOSchannel} where the DOS of our baseline device is reported at two locations along the CNT axis at $V_{GS}$ = 0.15 V and $V_{DS}$ = 0.5 V. The inclusion of e-e interactions significantly broadens the source valence band edge and increases its DOS by several orders of magnitude, while the channel DOS  remains relatively unchanged in the energy range of interest. This effect, captured by the imaginary part of the correlation self-energy $\Sigma^{R}_{corr}$ in Eq.~(\ref{eq:PI}) \cite{GWscreening}, comes from the fact that e-e scattering is much more pronounced in the source due to the higher carrier density there, leading to more interactions. The deeper the source DOS penetrates into the band gap, the higher the overlap between source and channel wave functions becomes, increasing tunnelling, OFF-state current, and SS. 

\section{Conclusions}
By including e-e interactions through the GW approximation into an \textit{ab initio} NEGF solver, we \CM{isolated} the role of Auger processes in the degradation of the sub-threshold swing of TFETs and confirmed the hypothesis presented in prior studies \cite{teherani2016auger,ahmed2020comprehensive}, namely the presence of e-e scattering opens up additional current channels, increasing the OFF-state current and the sub-threshold swing. Although the obtained SS does not exceed 60 mV/dec, \CM{it gets very close to it so that the inclusion of electron-phonon scattering, electron-ion interactions, and interface traps would push it above this limit. With the developed simulation approach, it might be possible to design novel device concepts that leverage Auger processes instead of suffering from them \cite{8421442}.}


\newpage

 \bibliographystyle{IEEEtran}

  \bibliography{tfet}


\end{document}